\documentclass[aps,twocolumn,prd,epsf]{revtex4}

\usepackage{graphicx}

\newcommand{\be}{\begin{equation}}
\newcommand{\ee}{\end{equation}}
\def\bq{\begin{eqnarray}}
\def\eq{\end{eqnarray}}
\def\beq{\begin{eqnarray}}
\def\eeq{\end{eqnarray}}

\def\ba{\begin{eqnarray}}
\def\ea{\end{eqnarray}}

\def\lra{\longrightarrow}

\newcommand{\mpl}{M_{\rm pl}}
\newcommand{\lpl}{\ell_{\rm pl}}

\begin{document}

\title{Big Crunch Avoidance in $\rm k = 1$ Semi-Classical 
 Loop Quantum Cosmology}      

\author{Parampreet Singh$^\ast$, Alexey Toporensky$^\dagger$}

\address{~}

\address{$^\ast$IUCAA, Ganeshkhind, Pune 411 007, India}
\email{param@iucaa.ernet.in}  

\address{$^\dagger$Sternberg Astronomical Institute, Moscow University, 
Moscow 119 899, Russia}
\email{lesha@sai.msu.ru}  
 
 
\begin{abstract}
It is well known that a closed universe with a minimally coupled massive
scalar field always collapses to a singularity unless the initial
conditions are extremely fine tuned. We show that the corrections to
the equations of motion for the massive scalar field,
 given by loop quantum gravity in high curvature
regime, always lead to a bounce independently of the initial conditions.
In contrast to the previous works in loop quantum cosmology,
we note that the singularity can be avoided even at the semi-classical
level of effective dynamical equations with non-perturbative 
quantum gravity modifications,
 without using a  discrete quantum evolution.
\end{abstract}
 
\pacs{pacs: 04.60.Kz, 98.80.Qc}
 
\maketitle

Various models in cosmology predict a dead end of our universe 
in a final state of collapse known as big crunch. It is well known that   
 a big crunch may occur if the universe
is matter dominated and starts collapsing under its own weight, leading to
a cosmic doomsday scenario.
The features of a classical cosmological evolution strongly depend on the
sign of spatial curvature. In standard cosmology a flat or open universe can  either experience
only  expansion or contraction without a transition from one regime
to another, unless it is filled by some exotic matter with a negative
energy density. Thus, if such a universe is expanding it never dies
in a big crunch.
On the contrary, a positive spatial curvature allows  transitions from
expansion to contraction and vice versa, moreover, in some models such
transitions are inescapable. With the current astronomical observations
 constraining $\Omega_0 = 1.02 \pm 0.02$ \cite{astro}, it is quite possible that our
universe may be closed and heading towards a big crunch. 
Apart from standard cosmology, big crunch scenario is also present
  in models of 
string cosmology \cite{superstring} and braneworlds \cite{kanti}.

In classical cosmology a $\rm k = 1$ universe if filled with a
matter source satisfying equation of state $\omega < -1/3$ 
(where $\omega = p/\rho$) could have
a finite minimum size, since in this case the
curvature term in Hubble equation dominates over the matter term 
for small $a$. Such a universe will neither have a big bang nor a 
big crunch. On the other hand if $\omega > -1/3$ then the universe starts
from a big bang and ends in a big crunch.
The problem of big crunch has been extensively studied in $\rm k = 1$
FRW model with a minimally coupled massive scalar field 
\cite{Starob,hawking,page,topo1,topo2,cornish}. 
Such a study captures the 
essential features of the dynamics before big crunch in general. 
Since,  the effective $\omega$ for a scalar field $\phi$ with 
some nonzero potential $V(\phi)$ can vary in the range $[-1,1]$, hence
 a variety
of possibilities exist. If $V(\phi)$ is a smooth function possessing a minimum
(this type of potential is favored by a theory of reheating),
then the scalar field at late epoch of cosmological expansion 
will mimic an ordinary matter with $\omega \ge 0$ \cite{turner} 
and the transition to a 
contracting phase proceeds. 
  We should also notice that a wide family of potentials
without a local minima invoked to explain issues of dark energy (see 
\cite{de1,de2,de3} for reviews)
lead to eternal expansion even in the case of $\rm k = 1$, and such models 
do not face the big crunch problem. This is also true for some potentials
with local minima as has been shown in \cite{roy} \& \cite{doran}.
In this work we will, however
only consider potentials with a local mimina in a framework which leads classically to a big crunch.

An important question which thus arises is whether a collapsing
universe with a scalar field experiences a bounce or ends in a 
big crunch. Answer to this depends not only on 
potential $V(\phi)$, but also on a particular phase trajectory chosen.  
Bounce is indeed possible for potentials which are not very steep,
but detailed investigations have shown,
that this in general requires a severe fine-tuning of initial conditions. 
Moreover,
the revealed chaoticity of such cosmological dynamics \cite{cornish}
indicates, that a bounce in the end of one cosmological cycle reflects
nothing about a possibility to have a bounce during the next contraction
stage of the universe. To ensure two consecutive bounces it is 
therefore 
necessary
to further constrain the initial conditions. As a 
result only a zero measure set of initial conditions gives us a possibility
to have infinite number of bounces, and then  to escape a singularity
for an infinite time interval. This leads to a conclusion that the future 
singularity
problem can not be solved in  classical cosmology with a massive scalar
field.

Occurrence of singularities in classical cosmology has always been
thought of as a signature of a domain
where general relativity ceases to be operational and must be replaced by
a quantum theory of gravity. In the present paper we  try to
understand dynamics leading to big crunch in  light of one of
the approaches to quantize general relativity known as loop quantum gravity.
It is based on canonical quantization of general relativity and its
salient 
features include background independence, 
non-perturbativeness and prediction of discrete spectrum for
geometrical operators \cite{loopqg}. This theory has  recently been
 applied in cosmological scenarios and has yielded
various novel results which include
 absence of big bang singularity \cite{martin1},
insights on initial conditions of the universe \cite{martin2},
possibility of inflation \cite{martin3} with right 
e-foldings \cite{tsr}, suppression of power at large scales in CMB \cite{tsr}
etc. For related works in this direction see for example \cite{all}.

One of the key features of loop quantum gravity is that it predicts
a modification in the behavior of geometrical density $a^{-3}$ 
at short distances.
Unlike  in the conventional quantum cosmology where the geometrical
density blows up when $a \lra 0$, in loop quantum cosmology the eigenvalues
of the corresponding operator become zero  at $a = 0$. We should note that near
$a \sim 0$ the picture of a continuous spacetime breaks down and
evolution is described by difference equations, however, there exists
a  domain where the continuous spacetime picture is valid and
 quantum gravity effects are manifest. 
 The eigenvalues of geometrical 
density operator in loop quantum cosmology are given as \cite{martin1},
\beq \label{dj}
d_j(a)=D(a^2/a_*^2)a^{-3}\,,~ a_*^2=\gamma \lpl^2\,j/3 \,,
\eeq
where $a$ is the scale factor, $\lpl$ is the Planck length,
$\gamma $ is the Barbero-Immirzi parameter and $j$ 
is a half-integer greater than unity. Here the function $D(a^2/a_*^2)$
is derived from quantum theory and is given by, 
 \ba \label{p}
D(q) &&= (8/77)^6 q^{3/2} \{7 [(q+1)^{11/4}
-|q-1|^{11/4}] \nonumber \\
&&{}- 11q[(q+1)^{7/4}-{\rm sgn}\,(q-1) |q-1|^{7/4}] \}^6\!, 
 \ea
where $q := a^2/a_*^2$. It should be noted that the scale where quantum 
effects become significant is defined by $a_*$ and though the Barbero-Immirzi
parameter can be fixed to $\gamma = \ln 2/(\sqrt{3} \pi)$ by black hole
thermodynamics, the parameter $j$ is arbitrary and can be used to set
an effective quantum gravity scale in the theory which may even be bigger than
$\lpl$.

The density operator in eq.(\ref{dj}) has various peculiar features.
In the quantum domain when $a \ll a_*$ it behaves as
\be
d_j \approx
(12/7)^6(a/a_*)^{15} a^{-3}\, .
\ee
In  transition to the classical regime when $a$ increases, $d_j$
effectively becomes proportional to lower powers of $a$ and eventually
becomes $d_j \approx a^{-3}$ when $a >> a_*$, which signals onset of
 a classical regime.

The Hamiltonian of a massive scalar field $\phi$ with a potential
$V(\phi)$ in loop quantum cosmology can be written as
 \be \label{hamiltonian}
{\cal H}= d_j \, \frac{p_\phi^2}{2} + a^3 \, V(\phi)\,,
\ee
where $p_\phi$ is the canonically conjugate momentum to $\phi$ and satisfies
$p_\phi = d_j^{-1}\dot \phi$. For the case of $\rm k = 1$ FRW model
the effective Friedmann equation becomes
\beq \label{fred_eq}
H^2 = \frac{8\pi}{3 \mpl^2} \,  \left( \frac{1}{D} \frac{\dot{\phi}^2}{2} + V(\phi) \right) \, - \frac{1}{a^2} 
 \eeq
together with the modified Klein-Gordon equation
 \beq
\ddot{\phi} &=& \nonumber a \, H\, \dot \phi \, \frac{d}{da} \ln d_j - a^3 \, d_j \, V_{,\phi} \\
&=& \left(-3 H + \frac{\dot D}{D} \right) \dot{\phi} - D \,  V_{,\phi} \,
\label{backp}
 \eeq
which yield the Raychaudhuri equation with quantum gravity correction as
\beq\label{rai}
\frac{\ddot a}{a} &=& \nonumber - \frac{2 \pi}{3 \mpl^2} \, \left(\frac{1}{D} \dot \phi^2 - \frac{1}{a^2}  \, \dot \phi^2 \frac{d}{da} d_j \right) + \frac{8 \pi}{3 \mpl^2} \, V(\phi) \\
&=& - \frac{8 \pi}{3 \mpl^2 D} \, \dot \phi^2 \, \left(1 - \frac{\dot D}{4 H D} \right) + \frac{8 \pi}{3 \mpl^2} \, V(\phi) ~.
\eeq

Before we study some interesting phenomenon due to quantum gravity 
modifications in above dynamical equations, it is important to emphasize
the semi-classical nature of these equations.
As stressed earlier, the picture of a continuous spacetime breaksdown
near $a \sim 0$, where the description is entirely in terms of spin network 
states. 
Hence, the dynamical equations (\ref{fred_eq},\ref{backp},\ref{rai}) are not valid
very close to $a \sim 0$. This  is the regime where full quantum gravity 
is functional and
the evolution can be described only in terms of quantum difference equations.
The dynamical equations are thus semi-classical in nature which are valid
in the domain around $a_*$, where we have a continuous spacetime but with
quantum gravity modifications.
 It is also interesting to note that
the full theory of loop quantum gravity does not have a notion of external
time which is present at the semi-classical level, so the notion of
trajectories which are solutions of the dynamical equations is invalid
near $a \sim 0$.  Thus if we wish to resolve the  initial big bang singularity then one has
to do a full quantum gravity treatment based on discrete form of the 
Wheeler-DeWitt equation \cite{martin1}. However, as we would show, to avoid
a big crunch singularity,  the modifications of dynamical equations at 
the above semi-classical level are sufficient.

Using eq.(\ref{dj}) it is easy to see that the $\dot \phi$ term  in Klein-Gordon
equation changes its sign as the universe evolves from a quantum regime
to a classical regime or vice versa. This behavior is crucial in establishing
various key results in loop quantum cosmology. For example, if the
universe is expanding from a quantum regime the usual friction term in
the Klein-Gordon equation is replaced by 
an anti-friction term which
is responsible for an inflaton to climb the potential hill, even if it starts
from its bottom, as has been
shown for the flat model \cite{martin3,tsr} and the close model \cite{ellis}.
It is important to note that the potential term in Klein-Gordon equation
becomes sub-dominant in quantum domain, since the function 
$D \ll 1$ for $a \ll a_*$.

If we use the conventional Klein-Gordon equation without
quantum correction in the case when 
universe is collapsing, then the $\dot \phi$ term in 
 classical regime acts like an anti-friction term
and the scalar field acquires very high values  just before big 
crunch. However, in loop quantum cosmology this situation will be 
prevented since when a collapsing universe reaches a size comparable 
or less than $a_*$, the otherwise anti-friction term becomes a frictional
term with a higher magnitude which stops the scalar field and 
leads to a bounce. In the quantum regime, the
strong dominance of the kinetic term
over the potential term means not only that the mechanism is
robust to a change in $V(\phi)$, but also that it dominates over
gradient terms in scalar field.

To demonstrate the effect of loop quantum gravity on big crunch let us
first consider the case of a quadratic potential
\be \label{quad}
V(\phi) = \frac{1}{2} \, m_\phi^2 \, \phi^2 ~.
\ee 
In the classical regime the equation of a massive scalar
field (\ref{backp}) describes the behavior of a harmonic oscillator.
 A detailed analysis of cosmological
solutions in this case can be found for example in \cite{Starob}. 
It is important to
note  that in the classical regime when the $\dot \phi$ term  is
anti-frictional during collapse,
almost all initial conditions lead to (possibly after
a period of growing scalar field oscillations) a regime where the 
potential term in Friedmann equation (\ref{fred_eq}) is 
 negligible in comparison to the kinetic term.
In this domain,
\begin{equation}
H=\frac{1}{3(t_0-t)}, \,  \qquad |\phi| ={\frac{\mpl}{\sqrt{12 \pi}}}\ln\frac{t_0-t}{c}
\end{equation}
and a singularity is reached at some moment $t=t_0$, where $t_0$ and
$c$ are constants of integration.
Some trajectories can avoid
this regime and experience bounce, giving rise to an interesting
fractal structure of the phase space (see \cite{page, topo1,cornish} for
details), but their measure is very small. Such a possibility
 is approximately inversely
proportional to $a_{max}/\lpl$, where $a_{max}$ is the scale factor 
at the point of maximal expansion \cite{Starob}. 

\begin{figure}[tbh!]
\begin{center}
\includegraphics[width=9cm,height=7.5cm]{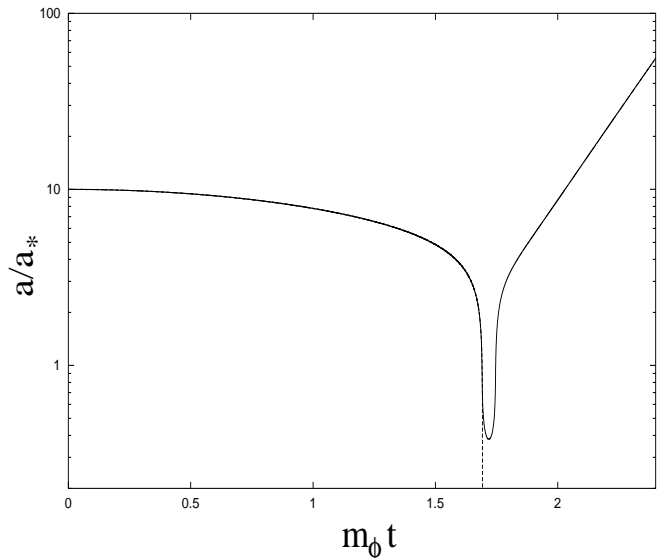}
\end{center}
\caption{Evolution of scale factor for a closed universe in classical
cosmology is shown by the dashed curve and  in loop
quantum cosmology is shown by the solid curve for the quadratic
potential with $j = 100$, $m_\phi = 0.1 \mpl$  and initial conditions
$a_i = 10 \, a_*, H_i = 0, \phi_i = 0$ and $ \dot \phi_i$ determined from
eq.(\ref{fred_eq}).  
A collapsing closed universe in classical cosmology encounters a 
big crunch whereas in loop quantum cosmology it bounces into
an expanding phase when $a = 0.38 \, a_*$ and avoids big crunch.} \label{fig1}
\end{figure}

\begin{figure}
\begin{center}
\includegraphics[width=9cm,height=7.5cm]{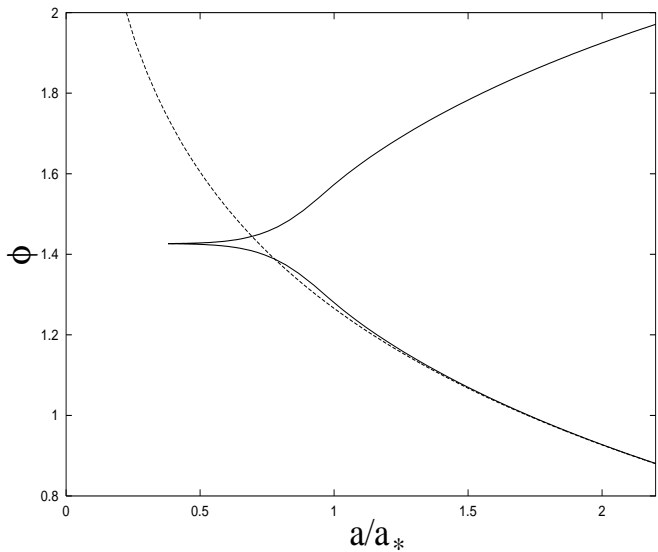}
\end{center}
\caption{Plot of $\phi$ in Planck units and $a/a_*$ in the domain where the
transition from classical to quantum and then again to classical occurs in
loop quantum cosmology for $\phi^2$ potential.
 The dashed curve is the one for classical
cosmology and the field approaches infinity as universe encounters
big crunch. The solid curve is for loop quantum cosmology evolution.
In this case it is seen that the field freezes near the bounce. Initial
conditions, $j$ and $m_\phi$ are same as in Fig. [\ref{fig1}].}\label{fig2}
\end{figure}

Solving eqs.(\ref{fred_eq} - \ref{rai}) along with eq.(\ref{dj})
numerically we can understand the change in dynamics due to quantum gravity
effects and as expected, the change from the classical equations is very
drastic. The evolution of scale factor for the quadratic potential is shown
in Fig. [\ref{fig1}]. We consider  initial conditions such that the
universe encounters a big crunch classically (as shown by a dashed curve). The evolution
of scalar field with respect to scale factor is shown in Fig. [\ref{fig2}].
As can be seen near the big crunch the field $\phi$ takes very large values.
It is clear from both of these figures that this situation is averted
in quantum gravity. 

When the size of the universe becomes 
comparable to $a_*$ and less, the quantum 
corrections become important in dynamical equations. As mentioned earlier, 
in the collapsing case the anti-friction term which was $- 3 H \dot \phi$ 
(with negative $H$) in classical regime, changes sign, becomes frictional and 
even goes as $12 H \dot \phi$ for $a \ll a_*$. In this process,
quantum gravity essentially applies breaks to the motion of $\phi$
through a large friction term in Klein-Gordon equation and
the scalar field almost completely
``freezes'' at some
high value $\phi_{f}$.
 At the instant
when the motion of $\phi$ ceases and $\dot \phi = 0$, all quantum gravity
signatures are excluded from the model and equation of state  becomes $-1$.
 Such equation of state in the case of positive spatial
curvature ultimately leads to a bounce. So, the contracting universe
bounces and turns to expansion.

After the bounce when the universe is still smaller than $a_*$, the $\dot \phi$
term in eq.(\ref{backp}) acts like an anti-friction term 
(since, the Hubble parameter $H$ changes sign to positive in expanding 
phase). The scale factor in this phase grows very rapidly and 
eventually as $a \gg a_*$, the quantum
corrections become negligible and the $\dot \phi$ term becomes
frictional.  The universe then
enters a standard inflationary stage. An example for this behavior
can be seen in variation of scale factor in Fig. [\ref{fig1}].
After checking a large range of initial conditions and parameters $j$ and 
$m_\phi$
we found that a bounce always occurs and the behavior of dynamics is
qualitatively the same as shown in Figs. [\ref{fig1} \& \ref{fig2}].

\begin{figure}[tbh!]
\begin{center}
\includegraphics[width=9cm,height=7.5cm]{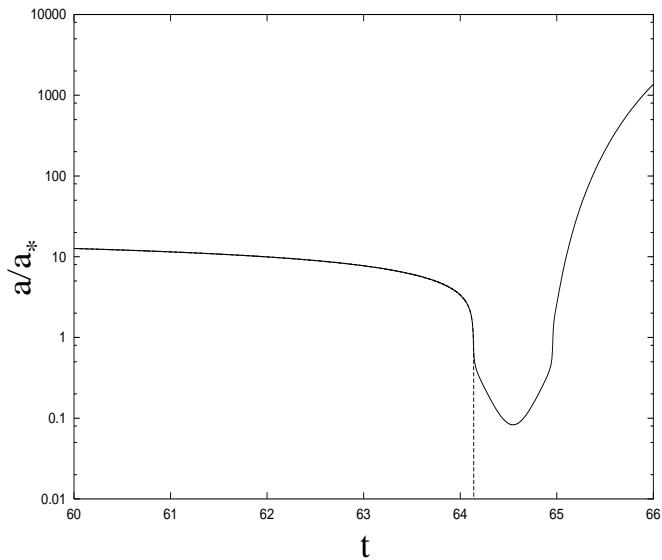}
\end{center}
\vskip-0.25cm
\caption{Scale factor for the case of steep potential (\ref{steep}) at late times (in Planck units) of
collapse. Initial conditions are $a_i = 100 \, a_*, H_i = 0$ and $\phi_i = 0$ (which also
determine $\dot \phi_i$)  
with $j = 100$. The evolution in loop quantum cosmology is depicted by solid
curve which avoid big crunch when $a = 0.082 a_*$ whereas classical
cosmology shown by dashed curve encounters a singularity.} \label{fig3}
\end{figure}

With the occurrence of bounce established for the quadratic potential in
a generic way we
now examine the case for a steep potential,
\be \label{steep}
V(\phi) = 2 \, \left(\cosh (\phi^2/\mpl^2) - 1 \right) ~.
\ee
The dynamics of a scalar field in classical cosmology is rather different
for potentials less and more steep than the exponential one. A steep potential
never becomes dynamically unimportant at a contraction stage, and a scalar
field oscillates infinitely while falling into a singularity \cite{Foster}.
The bounce for steep potentials as classically understood is impossible and
a thus a singularity is inevitable \cite{topo2}.

\begin{figure}[tbh!]
\begin{center}
\includegraphics[width=9cm,height=7.5cm]{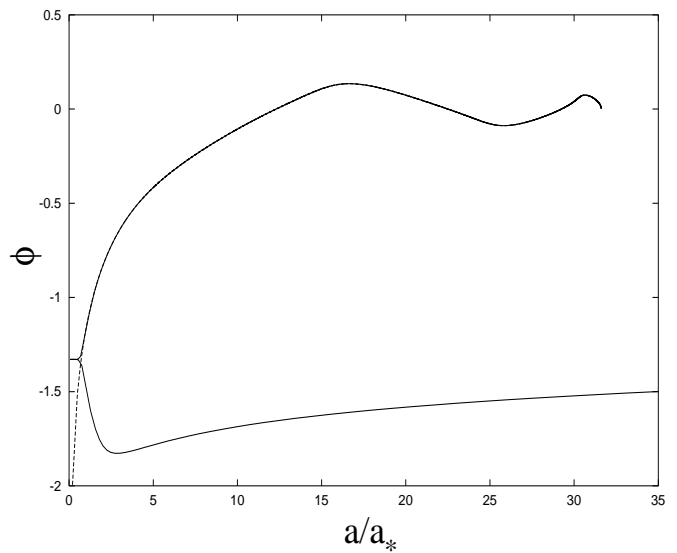}
\end{center}
\vskip-0.25cm
\caption{Dynamics of $\phi$ (in Planck units) and $a/a_*$ for potential 
(\ref{steep})
with initial conditions same as in Fig. [\ref{fig3}]. 
In the classical regime the field undergoes oscillations. In loop
quantum cosmology the field freezes near bounce (shown by solid curve) whereas
in classical cosmology (shown by dashed curve) it takes large values
near big crunch.} \label{fig4}
\end{figure}

These features, however, do not change the dynamics in quantum regime.
The quantum gravity corrections are significant mainly due to their 
influence on $\dot \phi$ terms in dynamical equations and thus in the quantum
regime they become dominant over the potential terms. The numerical results
for the potential (\ref{steep}) are shown in  Figs. [\ref{fig3} \&
\ref{fig4}] which clearly show a bounce.
The behavior of the universe near bounce is very similar to one shown in 
Figs. [\ref{fig1} \& \ref{fig2}].
In Fig. [\ref{fig4}] we have shown
how in classical regime the field $\phi$ undergoes oscillations and then
in quantum regime it first freezes and bounces back. As for the quadratic potential we found that the bounce for steep
potential (\ref{steep}) always occurs for arbitrary choice of
initial conditions and $j$. 

Our results are in a sharp contrast to earlier results in classical cosmology,
where a occurrence of a bounce requires
a very special initial condition to be 
chosen. As the quantum bounce is independent
of initial conditions, it eventually occurs also in the second, third and  
consecutive cycles 
of contractions giving rise to some kind of aperiodic cyclic universe.
It is interesting that the bounce via loop quantum gravity is rather similar 
to 
an artificial bounce put by hand into the classical picture in \cite{Nissim}
($\dot a \to -\dot a$ and $\dot \phi \to \dot \phi$ at some scale factor 
of the order of the Planck length). This suggests that  analysis of effects of
``hysteresis'' of the scalar field evolution 
 and time asymmetry which have been done in \cite{Nissim} are also
applicable to our model.

We should stress that for both considered potentials, quadratic (\ref{quad})
and steep (\ref{steep}), we found that the bounce occurs when size
of the universe is not very small compared to $a_*$. For the quadratic
potential a closed universe bounces when $a \sim 0.2 - 0.5 \, a_*$ and
for a steep potential it bounces when $a \sim 0.05 - 0.3 \, a_*$. 
We also
found that the role of $j$ on the nature of dynamics near big 
crunch is very weak. It is important to note that the big crunch singularity
has been avoided purely at the semi-classical level of effective
dynamical equations with quantum gravity modifications.
 This is in distinction to the earlier work in loop
quantum cosmology \cite{martin1}, 
where the singularity avoidance was accomplished 
using discrete quantum evolution near $a = 0$, in which case
 one can not use the 
picture of a continuous spacetime and the dynamical equations (\ref{fred_eq} - \ref{rai}) are invalid.

Thus, we establish that for a closed universe with a minimally coupled
massive scalar field  a bounce will occur for any choice of initial
conditions as soon as universe becomes of the size that its dynamics is
governed by equations with loop quantum gravity modifications. Hence a 
big crunch or a cosmic doomsday will  always be avoided.  
In future, it will be important to study dynamics of bounce in
anisotropic case and also the role of matter and radiation near bounce.
Apart from loop quantum cosmology, bouncing universe scenarios
 have been in discussion in various string models (for example see 
\cite{allstrings}). 
It will  be  interesting  to investigate their common features with
loop quantum effects 
for various values of geometrical curvature.

{\it Acknowledgements:} We thank M. Bojowald, N. Dadhich, R. Maartens,
M. Sami and S. Tsujikawa for extended discussions and useful comments
on this manuscript. PS thanks CSIR for a research grant and AT is supported
by IUCAA's `Program for enhanced interaction  with the Africa-Asia-Pacific region.' AT thanks IUCAA for hospitality where this work was carried out.

\end{document}